

\magnification=1200
\hsize=13cm
%
%
\centerline{\bf 1. Introduction and Conclusions}
\vskip 1cm
Quasitriangular Hopf algebra [1-3] has attracted a lot of attention of
physicists and mathematicians in the last few years [4]. Interesting
examples of this structure are deformations of Lie algebras and Lie
groups [1-4].

This structure has left its track in several areas of physics [5].
On the other hand, anyons [6-9], {\it i.e.}
two dimensional objects with
arbitrary statistics interpolating between bosons and fermions, have
also deserved considerable interest mainly because their possible
connection to planar condensed matter phenomena, particularly the
fractional quantum Hall effect [10].

Deformed Lie algebras and anyons, may have a deep connection as
indicated by the role played by the braid group in both these
structures, therefore it is possible that the characteristic symmetries
 of anyonic systems could be described by deformed Lie algebras.

It was shown in ref. [11] an interesting relation between these two
structures. It was constructed intrinsic two dimensional operators
on a two dimensional lattice, interpolating between fermionic and
bosonic oscillators.
These anyonic oscillators, which carry braiding properties, were used to
build explicitly the deformed algebra of $SU_q(2)$ in a sort of
generalized Schwinger construction [12].

Actually, a generalized Schwinger approach was built a few years ago,
using operators obeying $q$-commutation relations, in order to realize
the quantum enveloping algebra of the type $A_n~,~B_n~,~C_n~,~D_n$
[13-17] and the quantum exceptional algebras [18]. These
$q$-oscillators are different from the anyonic oscillators
because they can live in any dimension and are local operators.
 Therefore, it seems sensible to ask to what extent the role of the
building blocks of quantum algebras can be played by anyonic oscillators.

In this letter by considering a set
of $N$  anyonic oscillators we realize
 \`a la Schwinger  the  quantum enveloping
algebra of $SU_q(N)$. The main difference from the $SU_q(2)$ case
[11] comes from a trilinear relation among the quantum generators
called {\it deformed Serre's relation}, which is proved in section
{\bf 3} by non trivial use of the braiding properties of the
generator's density of $SU_q(N)$.

It comes out from [11] and from the case we have analysed to be
essential to deal with hard core anyonic oscillators to construct the
deformed algebras. Therefore we  believe that for all
deformed algebras which can be realized with $q$-deformed fermionic
oscillators the same role could be played by anyonic oscillators.

 In section {\bf 2} we review briefly the main results
concerning anyonic
oscillators, in section {\bf 3} we show how to realize the
 quantum enveloping algebra of $SU_q(N)$
with these oscillators and
 we find that the deformation parameter  is connected
to the anyonic statistical parameter $\nu$ as $q=\exp({\rm i}\pi\nu)$.

%
%
\vskip 4cm
\centerline{\bf 2. Lattice Angle Function and Anyonic Oscillators }
\vskip 1cm
In this section we are going to review the construction of anyonic
oscillators of ref. [11], on a two-dimensional square lattice of
lattice spacing one.

Anyonic oscillators are two-dimensional non-local objects [19-24],
interpolating between bosonic and fermionic oscillators which will be
constructed on a square lattice $\Omega$ by means of the Jordan-Wigner
transformation [25] that transmutes, in our case, fermionic oscillators
into anyonic oscillators. Its basic ingredients are a lattice angle
function and fermionic oscillators.

The lattice angle function was built in a very general way in ref.
[21], but we shall describe concisely the special but very useful way of
defining it of ref. [11].

We start by defining for each point ${\bf x}$ of the two dimensional
lattice $\Omega$ a cut ${\gamma}_x$, made with bonds of the dual lattice
$\tilde{\Omega}$ from minus infinity to ${\bf x}^{\ast}={\bf x}+{\bf o}
^{\ast}$ along $x$-axis,
with ${\bf o}^{\ast}=\left({1\over 2},{1\over 2} \right)$ the origin
of the dual lattice, and we denote by ${\bf x}_{\gamma}$ the point
${\bf x} \in \Omega $ with its associated cut ${\gamma}_x$.

Given any two distinct points ${\bf x}=(x_1,x_2)$ and
${\bf y}=(y_1,y_2)$ on $\Omega $,
and their associated cuts ${\gamma}_x$ and ${\gamma}_y$, it is possible
to show that [11]
$$
\Theta_{{\gamma}_x}({\bf x},{\bf y}) - \Theta_{{{\gamma}_y}}({\bf y},
{\bf x}) =
\left\{\matrix{
 \pi \,{\rm sgn}(x_2-y_2)  ~~~~~{\rm for}~~~x_2\not=y_2 \ \ , \cr
 \pi \,{\rm sgn}(x_1-y_1)  ~~~~~{\rm for}~~~x_2=y_2 \ \ ,
\cr}\right.
\eqno(2.1)
$$
with $\Theta_{{\gamma}_x}({\bf x},{\bf y})$ being
the angle of the point $\bf x$ measured from the point
${\bf y}^{\ast} \in {\tilde\Omega}$
with respect to a line parallel to the positive $ x$-axis.

In the formula (2.1) we are neglecting a term which depends
on ${\bf x}$ and ${\bf y}$ [21] and, as shown in ref.
 [11], as this term is a lattice feature this is done
by a sort of continuum limit. This term can be neglected
for all ${\bf x}, {\bf y} \in \Omega$.

We notice that the elimination of this term is
 essential to get $q$-commutation relations for anyonic
 oscillators with a constant $q$-factor as will be
clear in a moment.

The term depending on ${\bf x}$ and ${\bf y}$ which is missing
 in (2.1) can be seen [11] as being the angle between
${\bf x}$ and ${\bf x}+2{\bf o}^\ast$ as measured from
${\bf y}^\ast$, which is clearly negligible when
${\bf x}$ and ${\bf y}$ are very far apart from each other.
 However, this is not the case when ${\bf x}$ and ${\bf y}$
 are close to each other.

When ${\bf x}$ and ${\bf y}$ are close to each other we embed the
lattice $\Omega$ into another lattice $\Lambda$ whose lattice spacing
$\epsilon$ is taken to be much smaller than one. As $\Omega$ is
 a sublattice of $\Lambda$, the quantities defined on $\Omega$
can be viewed as the restriction to $\Omega$ of quantities defined on
$\Lambda$. In this case the missing term can be represented as the
angle between ${\bf x}$ and ${\bf x}+2\epsilon{\bf o}^\ast$
as measured from $\tilde{\bf y}^\ast={\bf y}+\epsilon{\bf o}^\ast$
which is negligible. Thus the missing term can be neglected
$\forall{\bf x},{\bf y}\in\Omega$ [11].

Equation (2.1), which relates the angle of two distinct
 points on $\Omega$, can be used to endow  the lattice with an ordering
which will be very useful
in handling anyonic oscillators. We define
${\bf x}_{\gamma} > {\bf y}_{\gamma} $
by making the choice of the positive sign
of equation (2.1), {\it i.e.}
$$
{\bf x}_{\gamma}>{\bf y}_{\gamma}
\Longleftrightarrow
\left\{\matrix{
 x_2>y_2 \ \ , \cr
 x_2=y_2~,~x_1>y_1 \ \ . \cr}\right.
\eqno(2.2)
$$

Even if it is unambiguous, this definition is not
unique, it depends on the choice made for the cuts.
Suppose, now, instead of choosing ${\gamma}_x$
we choose for each point of the lattice a cut ${\delta}_x$ made with
bonds of the dual lattice from, plus infinity to ${\ }^\ast{\bf x}$
along $x$-axis,
  with ${\ }^\ast{\bf x}={\bf x}-{\bf o}^{\ast}$ .
In this case it can be shown that the relation between
 the angle of two distinct points ${\bf x}, {\bf y} \in \Omega$
 becomes [11]
$$
\tilde\Theta_{{\delta}_x}({\bf x},{\bf y}) -
\tilde\Theta_{{{\delta}_y}}({\bf y},{\bf x})
=\left\{\matrix{
 -\pi \,{\rm sgn}(x_2-y_2)  ~~~~~{\rm for}~~~x_2\not
=y_2 \ \ , \cr
 -\pi \,{\rm sgn}(x_1-y_1)  ~~~~~{\rm for}~~~x_2
=y_2 \ \ . \cr}\right.
\eqno(2.3)
$$
Notice that $\tilde\Theta_{{\delta}_x}({\bf x},{\bf y})$
 is now the angle of {\bf x} as seen from ${\ }^\ast{\bf y} \in
{\tilde\Omega}$
with respect to a line parallel to the negative
$ x$-axis.

Denoting by ${\bf x}_\delta$ the point ${\bf x}$ with
its associated cut $\delta_x$ we can define
${\bf x}_{\delta} > {\bf y}_{\delta}$ for the positive
 sign of eq. (2.3) which in this case gives
$$
{\bf x}_{\delta}>{\bf y}_{\delta}
\Longleftrightarrow
\left\{\matrix{
 x_2<y_2 \ \ , \cr
 x_2=y_2~,~x_1<y_1 \ \ . \cr}\right.
\eqno(2.4)
$$
that is the opposite of the ordering induced by the cut $\gamma$.

We can also have the relation between $\Theta_{\gamma}$
and $\tilde\Theta_{\delta}$. Using the definition we have given
 for these angles we get [11]
$$
\tilde\Theta_{{\delta}_x}({\bf x},{\bf y}) -
\Theta_{{{\gamma}_x}}({\bf x},{\bf y})
=\left\{\matrix{
 -\pi \,{\rm sgn}(x_2-y_2)  ~~~~~{\rm for}~~~x_2\not
=y_2 \ \ , \cr
 -\pi \,{\rm sgn}(x_1-y_1)  ~~~~~{\rm for}~~~x_2=y_2
\ \ ,  \cr}\right.
\eqno(2.5)
$$
and using (2.3) and (2.5) it follows that
$$
\tilde\Theta_{{\delta}_x}({\bf x},{\bf y}) -
\Theta_{{{\gamma}_y}}({\bf y},{\bf x})
=0   ,
\eqno(2.6)
$$
which is valid also when ${\bf x}={\bf y}$, {\it i.e.}
$$
\tilde\Theta_{{\delta}_x}({\bf x},{\bf x}) -
\Theta_{{{\gamma}_x}}({\bf x},{\bf x})
=0   .
\eqno(2.7)
$$

We are going to use now, the two above angle functions
$\Theta_{{{\gamma}_x}}({\bf x},{\bf y})$ and
$\tilde\Theta_{{\delta}_x}({\bf x},{\bf y}) $
to define two kinds of parity related anyonic oscillators.

We define anyonic oscillators of type $\gamma$ and $\delta$
as
$$
a_{i}({\bf x}_{\alpha})=K_{i}({\bf x}_{\alpha})c_{i}({\bf x})
 ~~~~~~({\rm no~ sum~ over~} i)
\eqno(2.8)
$$
where the disorder operators $K_{i}({\bf x}_{\alpha})$
 [26] are given by
$$
\eqalign{
K_i({\bf x}_{\alpha}) &= {\rm e}^{{\rm i}\,
\nu\sum\limits_{{\bf y}\in \Omega}\,
\Theta_{{{\alpha}_x}}({\bf x},{\bf y})\,
c^\dagger_i({\bf y})c_i({\bf y})} }
\eqno(2.9)
$$
with $\alpha_{x} = \gamma_{x} , \delta_{x}  , ~~~~i=1,\cdots,N$,
{}~~ $\nu$ is the statistical parameter and $c_i$,$c^\dagger_i$
are fermionic oscillators defined on $\Omega$ which have the
well-known anticommutation relations
$$
\eqalign{
\Big\{c_i({\bf x})~,~c^\dagger_j({\bf y})\Big\}&=\,
\delta_{i,j}\delta({\bf x}{\bf y})\,
\ \,\cr
 \Big\{c_i({\bf x})~,~c_j({\bf y})\Big\}&=0\ \,\cr
\Big\{c^\dagger_i({\bf x})~,~c^\dagger_j({\bf y})\Big\}&=0
}
\eqno(2.10)
$$
with $\delta({\bf x},{\bf y})$ the delta function on $\Omega$,
 {\it i.e.} zero for different points of the lattice and
one if ${\bf x}={\bf y}$.

Using (2.1) and (2.10) we get for  type $\gamma$
 anyonic oscillators the following commutation relations
$$
\eqalignno{
a_i({\bf x}_{\gamma})~a_i({\bf y}_{\gamma})&+\,
q^{-1}~a_i({\bf y}_{\gamma})~a_i({\bf x}_{\gamma})=0 \ \ ,
&(2.11{\rm a})\cr
a_i({\bf x}_{\gamma})~a^\dagger_i({\bf y}_{\gamma})&+\,
q~a^\dagger_i({\bf y}_{\gamma})~a_i({\bf x}_{\gamma})=0 \ \ ,
&(2.11{\rm b})
}
$$
for ${\bf x}>{\bf y}$ (from now on ${\bf x}>{\bf y}
\Longleftrightarrow {\bf x}_\gamma>{\bf y}_\gamma$)
 and $q=\exp({\rm i}\pi\nu)$.
If ${\bf x}={\bf y}$ we have
$$
\left(a_i({\bf x}_\gamma)\right)^2=0 ,
\eqno(2.12{\rm a})
$$
$$
a_i({\bf x}_{\gamma})~a^\dagger_i({\bf x}_{\gamma})+
a^\dagger_i({\bf x}_{\gamma})~a_i({\bf x}_{\gamma})=1 .
\eqno(2.12{\rm b})
$$
Eqs. (2.11-12) tell us that these hard core oscillators
obey $q$-commutation relations for different points of the lattice
but the standard anticommutation relations for oscillators at the
same point of the lattice. We notice that we can get the complete
set  of relations for these oscillators by taking the hermitean
conjugate of formulas (2.11-12).

It is very easy to see that different oscillators obey the well
known anticommutation relations
$$
\eqalign{
\Big\{a_i({\bf x}_\gamma)~,~a_j({\bf y}_\gamma)\Big\}=
\Big\{a_i({\bf x}_\gamma)~,~a^\dagger_j({\bf y}_\gamma)\Big\}=0\cr
\forall~ {\bf x},\,{\bf y}\in\Omega ~~~~{\rm and}
{}~~\forall i,j~,~~i\not=j=1,\cdots,N .
}
\eqno(2.13)
$$

The commutation relations among anyonic oscillators of
type $\delta$ can be obtained from the previous ones,
(2.11-13), by replacing $q$ by $q^{-1}$ and $\gamma$
by $\delta$. This is due to the fact that
${\bf x}>{\bf y}$ means ${\bf x}_\gamma>{\bf y}_\gamma$
and thus ${\bf x}_\delta<{\bf y}_\delta$; alternatively
we can say that $\delta$ ordering can be obtained from
$\gamma$ ordering by a parity transformation which,
 as is well known, changes the braiding phase $q$ into
$q^{-1}$ (see for istance [9]).

To complete our discussion on the commutation relations of anyonic
oscillators we can compute these relations among different types of
oscillators, {\it i.e.} between type $\gamma$ and type $\delta$
oscillators. By using  eqs. (2.5-8) we have
$$
\eqalignno{
\Big\{a_i({\bf x}_{\delta})~,~a_i({\bf y}_{\gamma})\Big\}&=\,
0~~~~~~~~~~~\forall ~{\bf x},\,{\bf y} \ \ ,
&(2.14{\rm a})\cr
\Big\{a_i({\bf x}_{\delta})~,~a^\dagger_i({\bf y}_{\gamma})\Big\}&=\,
0~~~~~~~~~~~\forall ~{\bf x},\,{\bf y}~~~~{\bf x}\not={\bf y} \ \ ,
&(2.14{\rm b})
}
$$
and finally
$$
\Big\{a_i({\bf x}_{\delta})~,
{}~a^\dagger_i({\bf x}_{\gamma})\Big\}=
q^{-\Big[\sum\limits_{{\bf y}<{\bf x}}-
\sum\limits_{{\bf y}>{\bf x}}\Big]
c^\dagger_i({\bf y})\,c_i({\bf y})} \ \
\eqno(2.15)
$$
and those coming from taking the hermitean conjugate of relations
 (2.14-15).

It should be clear from the above discussion that anyonic
 oscillators does not have anything to do with $q$-oscillators
 introduced a few years ago (ref. [13-18]). The main reason is,
as we have seen, that anyonic oscillators depend on the introduction
 of an angle function, with its associate cut, which is an intrinsic
 two-dimensional object thus, as it should be, these oscillators are non
-local objects which can be introduced only in two dimensions,
contrarily
to the deformed $q$-oscillators which are local objects that can be
defined in any dimension.

%
%
\vfill
\eject
\centerline{\bf 3. Realization of ${\bf SU_q(N)}$ Algebra}
\vskip 1cm

We are going to realize, in this section, the quantum
enveloping algebra  $SU_q(N)$
 [1,2] with the anyonic oscillators defined in the previous section.

Let $\alpha_1,\cdots,\alpha_{N-1}$ be
the simple roots of the simple Lie
algebra $SU(N)$ and $a_{ij}$, for $i,j=1,\cdots,N-1$ ,
 its Cartan matrix.
The quantum enveloping algebra $SU_q(N)$ is a Hopf algebra with unit
$\bf 1$ and generators $E_{\pm \alpha_i}~,~H_{\alpha_i}$
defined through the commutation relations in the Chevalley basis
$$
\eqalignno{
\Big[H_{\alpha_i}~,~H_{\alpha_j}\Big]&=\,
0 \ \ ,
&(3.1{\rm a})\cr
\Big[H_{\alpha_i}~,~E_{\pm\alpha_j}\Big]&=\,
\pm a_{ij}E_{\pm\alpha_j} \ \ ,
&(3.1{\rm b})\cr
\Big[E_{\alpha_i}~,~E_{-\alpha_j}\Big]&=\,
\delta_{ij}{{q^{H_i}-q^{-H_i}}\over{q-q^{-1}}} \ \ ,
&(3.1{\rm c})\cr
\Big[E_{\alpha_i}~,~E_{\alpha_j}\Big]&=\,
0~~~~~~{\rm if}~~a_{ij}=0 \ \ ,
&(3.1{\rm d})
}
$$
and the deformed Serre's relation
$$
\sum_{n=0}^{1-a_{ij}}(-1)^n{{1-a_{ij}}\brack n}_q
\left(E_{\pm\alpha_i}\right)^{1-a_{ij}-n}E_{\pm\alpha_j}
\left(E_{\pm\alpha_i}\right)^n=0 \ \
\eqno(3.2)
$$
for $j=i+1$, with the $q$-binomial coefficients
$
{{m\brack n}}_q
$,
given by
$$
{{m\brack n}}_q={{{[m]}_q!}\over {{[m-n]}_q!~{[n]}_q!}}
\eqno(3.3)
$$
where
$
{[m]}_q!={[m]}_q{[m-1]}_q\cdots{[1]}_q
$,
and
$$
{[m]}_q={{q^m-q^{-m}}\over{q-q^{-1}}}. \ \
\eqno(3.4)
$$
To complete the definition the comultiplication
$\Delta$
 and antipode $S$ are given by
$$
\eqalign{
\Delta(H_{\alpha_i}) &=\,
 H_{\alpha_i}\otimes{\bf 1}+{\bf 1}\otimes H_{\alpha_i}\ \ ,
\cr
\Delta(E_{\pm\alpha_i})&=\,
E_{\pm\alpha_i}\otimes q^{-{H_{\alpha_i}}/2}+
q^{{H_{\alpha_i}}/2}\otimes E_{\pm\alpha_i}\ \ ,
\cr
S(H_{\alpha_i})&=\,
-H_{\alpha_i}\ \ ,
\cr
S(E_{\pm\alpha_i})&=\,
q^{-\rho}E_{\pm\alpha_i}q^\rho ,
{}~~~\rho={1\over2}\sum_{\alpha\in\Delta_+} H_{\alpha}
}
\eqno(3.5)
$$
with $\Delta_+$ the set of positive roots of $SU(N)$, and the co-unit
$$
\epsilon({\bf 1})=1~ ,~~~~~
\epsilon(E_{\pm\alpha_i})=\epsilon(H_{\alpha_i})=0 .
\eqno(3.6)
$$

Now, we realize the generators of $SU_q(N)$ quantum enveloping algebra
 through the $N$ anyonic oscillators, defined on the lattice, of the
previous section. Generalizing the $SU_q(2)$ case [11] the generators
should be given by the sum over the lattice of density operators.
Since type $\gamma$ and type $\delta$ oscillators are related by a
parity transformation which changes $q$ into $q^{-1}$, and as we expect
that
$$
E_{+\alpha_i}({\bf x})_q={\Big[E_{-\alpha_i}
({\bf x})_{q^{-1}}\Big]}^\dagger ,
\eqno(3.7)
$$
for $q$ complex, we are led to constuct $E_{\alpha_i}({\bf x})$
 and $E_{-\alpha_i}({\bf x})$ with different type of anyonic oscillators.

The density of quantum generators are given by
$$
\eqalignno{
E_{\alpha_i}({\bf x})&=\,
a^\dagger_i({\bf x}_\gamma)~a_{i+1}({\bf x}_\gamma)\ \ ,
&(3.8{\rm a})
\cr
E_{-\alpha_i}({\bf x})&=\,
a^\dagger_{i+1}({\bf x}_\delta)~ a_i({\bf x}_\delta)\ \ ,
&(3.8{\rm b})
\cr
H_{\alpha_i}({\bf x})&=\,
a^\dagger_i({\bf x}_\gamma)~
a_i({\bf x}_\gamma)-a^\dagger_{i+1}({\bf x}_\gamma)~
a_{i+1}({\bf x}_\gamma)=\cr
&=a^\dagger_i({\bf x}_\delta)~ a_i({\bf x}_\delta)-
a^\dagger_{i+1}({\bf x}_\delta)~ a_{i+1}({\bf x}_\delta)\ \ ,
&(3.8{\rm c})
}
$$
the last identity in eq. (3.8 c) comes from the cancellation
 of the disorder operators, as it may be easily checked using eq.
(2.8), and the generators are given by a sum on the lattice
$$
\eqalign{
E_{\pm\alpha_i} &=\,
\sum_{{\bf x}\in \Omega}E_{\pm\alpha_i}({\bf x})
\cr
H_{\alpha_i} &=\sum_{{\bf x}\in \Omega} H_{\alpha_i}({\bf x}).
}
\eqno(3.9)
$$

We can easily see that the equations (3.1 a) and (3.1 b) are
satisfied if we realize that $H_{\alpha_i}$ counts the number of
operators on the lattice of the type $i$ minus those of type
$i+1$, independently if they are type $\gamma$ or type $\delta$
operators. The relation (3.1 a) is really trivial, and separating eq.
(3.1 b) in four cases where
$\Big[H_{\alpha_i}~,~E_{\pm\alpha_i}\Big]$,
$\Big[H_{\alpha_i}~,~E_{\pm\alpha_{i+1}}\Big]$,
$\Big[H_{\alpha_i}~,~E_{\pm\alpha_{i-1}}\Big]$,
$\Big[H_{\alpha_i}~,~E_{\pm\alpha_j}\Big]$ for
$|i-j|\geq2 $ , we immediately realize that we get respectively $\pm2~,~
\mp1~,~\mp1$ and zero, which is the desired result.

It is again easy to get convinced that eq. (3.1 d) is satisfied. The
Cartan matrix $a_{ij}$ is equal to zero for $|i-j|\geq2$ and in this
case the generators $E_{\alpha_i}$ and $E_{\alpha_j}$ are made up with
different anyonic oscillators, which anticommute (eq. (2.13)),
giving thus the correct result.

The proof of (3.1 c) is analogous to the $SU_q(2)$ case [11].
We consider firstly the commutator of the generator's density
$E_{\alpha_i}({\bf x})$ and $E_{-\alpha_i}({\bf y})$, which gives zero
if ${\bf x}\not={\bf y}$, and using (2.14-15) we have
$$
\eqalign{
\left[ E_{\alpha_i}({\bf x})~,~E_{-\alpha_i}({\bf x})\right] =&
\,q^{\Big[\sum\limits_{{\bf y}<{\bf x}}-\,
\sum\limits_{{\bf y}>{\bf x}}\Big]c^\dagger_{i+1}({\bf y})\,
c_{i+1}({\bf y})}~a^\dagger_i({\bf x}_{\gamma})\,
a_i({\bf x}_{\delta})
\cr
&-q^{-\Big[\sum\limits_{{\bf y}<{\bf x}}-\,
\sum\limits_{{\bf y}>{\bf x}}\Big]c^\dagger_i({\bf y})\,
c_i({\bf y})}~a^\dagger_{i+1}({\bf x}_{\delta})\,
a_{i+1}({\bf x}_{\gamma})
\ \ .}
\eqno(3.10)
$$
Substituting the explicit definition of the anyonic oscillators
 in the right hand side, using (2.5) and (2.7) and taking into
 account that the above commutator is zero for
${\bf x}\not={\bf y}$, we obtain
$$
\left[E_{\alpha_i}~,~E_{-\alpha_i}\right] =
\sum_{{\bf x}\in \Omega}
\left(\prod_{{\bf y}<{\bf x}}q^{-H_{\alpha_i}({\bf y})}~
H_{\alpha_i}({\bf x})~
\prod_{{\bf z}>{\bf x}}q^{H_{\alpha_i}({\bf z})}\right) .
\eqno(3.11)
$$
Analogously to the $SU_q(2)$ case, by considering that for
 each point of the lattice $\Omega$ $H_{\alpha_i}({\bf x})$
admits only the eigenvalues $0$ and $\pm {1\over2}$, we can prove by
complete induction that the right hand side of (3.11) is
 indeed the right hand side of (3.1 c) for $i=j$. We would like
 to stress here that it is essential to choose hard core
oscillators to get this result. When $i$ is different from $j$,
let us consider $j$ as given by $|i-j|\geq1$. For the equal sign case,
 $E_{\alpha_i}$ and $E_{-\alpha_{i\pm 1}}$ share only
$a_{i+1}({\bf x}_\gamma)$  and  $a_{i+1}({\bf y}_\delta)$ or
$a^{\dagger}_{i}({\bf x}_\gamma)$ and $a^\dagger_{i}({\bf y}_\delta)$
respectively  as oscillators
of the same type, all the others are of different types. But from
(2.14 a) these oscillators anticommute $\forall {\bf x},{\bf y}$
thus $E_{\alpha_i}$ commutes with $E_{-\alpha_{i\pm 1}}$. When
$|i-j|>1$ all the oscillators which are present are of different kind
(anticommute among themselves) and as a result $E_{\alpha_i}$
commutes trivially with $E_{\alpha_j}$. Equation (3.1 c) is proved.

The deformed Serre's relation can be written explicitly for the case we
are analising as
$$
\left(E_{\pm\alpha_i}\right)^2~E_{\pm\alpha_{i+1}}-\left(q+q^{-1}\right)
{}~E_{\pm\alpha_i}~E_{\pm\alpha_{i+1}}~E_{\pm\alpha_i}+
E_{\pm\alpha_{i+1}}~\left(E_{\pm\alpha_i}\right)^2=0  .
\eqno(3.12)
$$

To prove (3.12) we have to write the generators in the above formula
in terms of the generator's density by the use of (3.9).
Our strategy will be that of dividing the sum over the lattice
into two parts, the sum where the generator's density are at different
lattice points and that one where at least two of the generator's
density are at the same lattice point. Each part can be compared
separately in formula (3.12).

Let us consider the first part. The first term of (3.12) can be written
as
$$
\eqalign{
I_1&=\sum_{{\bf x}\not={\bf y}\not={\bf z}}E_{\alpha_i}({\bf x})
E_{\alpha_i}({\bf y})E_{\alpha_{i+1}}({\bf z})=\cr
&\left(\sum_{{{\bf x}\not={\bf y}\not={\bf z}}\atop{{\bf y}>{\bf z}}}+
\sum_{{{\bf x}\not={\bf y}\not={\bf z}}\atop{{\bf y}<{\bf z}}}\right)
E_{\alpha_i}({\bf x})
E_{\alpha_i}({\bf y})E_{\alpha_{i+1}}({\bf z}) ;
}
\eqno (3.13)
$$
now using
$$
E_{\alpha_i}({\bf x})E_{\alpha_{i+1}}({\bf y})=\left\{\matrix{
 q~~ E_{\alpha_{i+1}}({\bf y})E_{\alpha_i}({\bf x})  ~~~~~{\rm for}~~~
{\bf x}>{\bf y}\ \ , \cr
 q^{-1}E_{\alpha_{i+1}}({\bf y})E_{\alpha_i}({\bf x})
  ~~~~~{\rm for}~~~{\bf x}<{\bf y}\ \ , \cr}\right.
\eqno(3.14)
$$
we have
$$
I_1=\left(q\sum_{{{\bf x}\not={\bf y}\not={\bf z}}\atop
{{\bf y}>{\bf z}}}+q^{-1}
\sum_{{{\bf x}\not={\bf y}\not={\bf z}}\atop{{\bf y}<{\bf z}}}\right)
E_{\alpha_i}({\bf x})
E_{\alpha_{i+1}}({\bf z})E_{\alpha_i}({\bf y}) .
\eqno(3.15)
$$
Using the same procedure for the third term of (3.12) we get
$$
\eqalign{
I_2&=\sum_{{\bf x}\not={\bf y}\not={\bf z}}E_{\alpha_{i+1}}({\bf z})
E_{\alpha_i}({\bf x})E_{\alpha_i}({\bf y})=\cr
&\left(q\sum_{{{\bf x}\not={\bf y}\not={\bf z}}\atop
{{\bf z}>{\bf x}}}+q^{-1}
\sum_{{{\bf x}\not={\bf y}\not={\bf z}}\atop{{\bf z}<{\bf x}}}\right)
E_{\alpha_i}({\bf x})
E_{\alpha_{i+1}}({\bf z})E_{\alpha_i}({\bf y})
}
\eqno (3.16)
$$
and it is easy to show using (3.14) and
$$
E_{\alpha_i}({\bf x})E_{\alpha_i}({\bf y})=\left\{\matrix{
 q^{-2}~ E_{\alpha_i}({\bf y})E_{\alpha_i}({\bf x})
{}~~~~~{\rm for}~~~{\bf x}>{\bf y}\ \ , \cr
 q^2~~E_{\alpha_i}({\bf y})E_{\alpha_i}({\bf x})
{}~~~~~{\rm for}~~~{\bf x}<{\bf y}\ \ . \cr}\right.
\eqno(3.17)
$$
that
$$
\sum_{{\bf x}<{\bf z}<{\bf y}}E_{\alpha_i}({\bf x})
E_{\alpha_{i+1}}({\bf z})E_{\alpha_i}({\bf y})=
\sum_{{\bf y}<{\bf z}<{\bf x}}E_{\alpha_i}({\bf x})
E_{\alpha_{i+1}}({\bf z})E_{\alpha_i}({\bf y}) .
\eqno(3.18)
$$

In the right hand side of (3.16) there is a term proportional to $q$
which is the sum over ${\bf x}<{\bf z}<{\bf y}$ that is already
present in (3.15), but
from (3.18) this term is equal to the sum over
${\bf y}<{\bf z}<{\bf x}$. The same happens
in (3.16) for the term proportional to $q^{-1}$ where the sum over
${\bf y}<{\bf z}<{\bf x}$ is already in (3.15), but again from (3.18)
this sum is equal to
the one over ${\bf x}<{\bf z}<{\bf y}$. Taking into account these two
observations we
realize that (3.16) is the missing part in (3.15) to make the opposite
of the second term in (3.12), {\it i.e.}
$$
\eqalign{
&\sum_{{\bf x}\not={\bf y}\not={\bf z}} \Big(E_{\alpha_i}({\bf x})\,
E_{\alpha_i}({\bf y})E_{\alpha_{i+1}}({\bf z})+\,
E_{\alpha_{i+1}}({\bf z})\,
E_{\alpha_i}({\bf x})E_{\alpha_i}({\bf y})\Big)=\cr
&= \Big(q+q^{-1}\Big)\,
\sum_{{\bf x}\not={\bf y}\not={\bf z}}E_{\alpha_i}({\bf x})\,
E_{\alpha_{i+1}}({\bf z})E_{\alpha_i}({\bf y})
\ \ .}
\eqno(3.19)
$$

To complete the proof we have to consider the part of the sum where at
least two of the generator's density are at the same lattice point. The
easiest way to handle this part is never cross generator's density at
the same lattice point.
The computation goes as follows. We first define the simbol
$\sum_{{\bf x},{\bf y},{\bf z}}^e$ as denoting the sum over
${\bf x},{\bf y},{\bf z}$ when at least two of the lattice points are
equal; notice that in our case we cannot have all of them being equal
because $\left(E_{\alpha_i}({\bf x})\right)^2=0$. Writing explicitly
this definition for the first term of (3.12) and using the braiding
relations (3.15) and (3.17) we have
$$
\eqalign{
&\sum_{{\bf x},{\bf y},{\bf z}}^e~E_{\alpha_i}({\bf x})E_{\alpha_i}\,
({\bf y})E_{\alpha_{i+1}}({\bf z})\equiv\cr
&\equiv\left(\sum_{{\bf y}<{\bf x}={\bf z}}+\,
\sum_{{\bf y}>{\bf x}={\bf z}}\right)\,
\left(E_{\alpha_i}({\bf z})\,
E_{\alpha_i}({\bf y})E_{\alpha_{i+1}}({\bf z})\right)+\cr
&+\left(\sum_{{\bf x}<{\bf y}={\bf z}}+\,
\sum_{{\bf x}>{\bf y}={\bf z}}\right)\,
\left(E_{\alpha_i}({\bf x})\,
E_{\alpha_i}({\bf z})E_{\alpha_{i+1}}({\bf z})\right)=
\cr
&=\left(q^{-1}\sum_{{\bf y}<{\bf x}={\bf z}}+\,
q\sum_{{\bf y}>{\bf x}={\bf z}}+\,
q\sum_{{\bf y}<{\bf x}={\bf z}}+\,
q^{-1}\sum_{{\bf y}>{\bf x}={\bf z}}\right)\,
\left(E_{\alpha_i}({\bf x})\,
E_{\alpha_{i+1}}({\bf z})E_{\alpha_i}({\bf y})\right)
\ \ .}
\eqno(3.20)
$$

Repeating the above steps for the third term of (3.12) we realize,
again, that this term is what is missing in (3.20) to make the opposite
of the second term, for the same kind of sum, in (3.12), {\it i.e.}
$$
\eqalign{
&\sum_{{\bf x},{\bf y},{\bf z}}^e~\Big(E_{\alpha_i}({\bf x})\,
E_{\alpha_i}({\bf y})E_{\alpha_{i+1}}({\bf z})+\,
E_{\alpha_{i+1}}({\bf z})\,
E_{\alpha_i}({\bf x})E_{\alpha_i}({\bf y})\Big)=\cr
&=\Big(q+q^{-1}\Big)\,
\sum_{{\bf x},{\bf y},{\bf z}}^e~E_{\alpha_i}({\bf x})\,
E_{\alpha_{i+1}}({\bf z})E_{\alpha_i}({\bf y})
\ \ .}
\eqno(3.21)
$$
Putting together (3.18) and (3.21) we get the desired result. The part
with minus sign in (3.12) is trivially obtained if we recall that under
a parity transformation $E_{\alpha_i}\rightarrow E_{-\alpha_i}$ and
$q\leftrightarrow q^{-1}$, and that (3.12) is invariant under
 $q\leftrightarrow q^{-1}$.

%
%
\vskip 3cm
\centerline{\bf Acknowledgments}
\vskip 1cm
\noindent
The authors would like to thank S. Sciuto for his encouragement,
interest and for careful reading the manuscript.

%
%
\vfill
\eject
\centerline{\bf References}
\vskip 1cm
\item{[1]}V.G. Drinfeld, {\it Sov. Math. Dokl.} {\bf 32}
(1985) 254;
\medskip
\item{[2]}M. Jimbo, {\it Lett. Math. Phys.} {\bf 10}
(1985) 63; {\bf 11} (1986) 247;
\medskip
\item{[3]} L.D. Faddeev, N. Yu. Reshetikhin and
 L.A. Takhtadzhyan , {\it Algebra and Analysis} {\bf 1} (1987) 178;
\medskip
\item{[4]}For reviews see for example: S. Majid,
{\it Int. J. Mod. Phys.} {\bf A5} (1990) 1; S. Sciuto, Lectures given at
the meeting {\it "Condensed Matter and High Energy Physics"} in Chia
(1990)- Cagliari- Italy , edited by L. Alvarez-Gaum\'e and S. Fubini;
 P. Aschieri and L. Castellani, {\it "An Introduction to
Non-Commutative Differential Geometry on Quantum Groups"}, Preprint
CERN-Th 6565/92, DFTT-22/92 to appear in {\it Int. J. Mod. Phys};
\item{[5]}C. Zachos,{\it "Paradigms of Quantum Algebras"},
Preprint ANL-HEP-PR-90-61;
\medskip
\item{[6]}J.M. Leinaas and J. Myrheim, {\it Nuovo Cim.}
{\bf 37B} (1977) 1;
\medskip
\item{[7]}F. Wilczek, {\it Phys. Rev. Lett.}
{\bf 48} (1982) 114;
\medskip
\item{[8]}F. Wilczek, in {\it Fractional Statistics
and Anyon Superconductivity} edited by F. Wilczek
(World Scientific Publishing Co., Singapore 1990);
\medskip
\item{[9]}For a review see for example:
A. Lerda, {\it Anyons: Quantum Mechanics of Particles
with Fractional Statistics} (Springer-Verlag, Berlin, Germany 1992);
\medskip
\item{[10]}For a review see for example: {\it The Quantum Hall
Effect} edited by R.E. Prange and S.M. Girvin (Springer-Verlag,
Berlin, Germany 1990);
\medskip
\item{[11]}A. Lerda and S. Sciuto, {\it "Anyons and Quantum Groups"},
 Preprint DFTT 73/92, ITP-SB-92-73;
\medskip
\item{[12]}J. Schwinger, in {\it Quantum Theory of
Angular Momentum} edited by L.C. Biedenharn and
H. Van Dam (Academic Press, New York, NY, USA 1965);
\medskip
\item{[13]}A. Macfarlane, {\it J. Phys.} {\bf A22}
(1989) 4581;
\medskip
\item{[14]}L.C. Biedenharn, {\it J. Phys.} {\bf A22}
(1989) L873;
\medskip
\item{[15]}T. Hayashi, {\it Comm. Math. Phys.} {\bf 127}
(1990) 129;
\medskip
\item{[16]}O.W. Greenberg, {\it Phys. Rev. Lett.} {\bf 64} (1990) 705;
\medskip
\item{[17]}C-P. Sun and H-C. Fu,  {\it J. Phys.}
 {\bf A22} (1989) L983;
\medskip
\item{[18]}L. Frappat, P. Sorba and A. Sciarrino, {\it J. Phys.}
 {\bf A24} (1991) L179;
N. Y. Reshetikhin, {\it "Quantified Universal Enveloping Algebras, the
Yang-Baxter Equation and Invariant of Links"}, Preprint II LOMI (1988);
\medskip
\item{[19]} E. Fradkin, {\it Phys. Rev. Lett.}
{\bf 63} (1989) 322;
\medskip
\item{[20]}M. L\"{u}scher, {\it Nucl. Phys.}
{\bf B326} (1989) 557;
\medskip
\item{[21]}V.F. M\"{u}ller, {\it Z. Phys.}
{\bf C47} (1990) 301;
\medskip
\item{[22]}D. Eliezer and G.W. Semenoff,
{\it Phys. Lett.} {\bf B266} (1991) 375;
\medskip
\item{[23]}D. Eliezer, G.W. Semenoff and S.S.C. Wu,
{\it Mod. Phys. Lett.} {\bf A7} (1992) 513;
\medskip
\item{[24]}D. Eliezer and G.W. Semenoff,
{\it Ann. Phys.} {\bf 217} (1992) 66;
\medskip
\item{[25]}P. Jordan and E.P. Wigner, {\it Z. Phys.}
{\bf 47} (1928) 631;
\medskip
\item{[26]}E. Fradkin, {\it Field Theories of
Condensed Matter Systems} (Addison-Wesley, Reading, MA, USA 1991).
\medskip

%
%
\vfill
\eject

\nopagenumbers
\hskip 9cm \vbox{\hbox{DFTT 5/93}
\hbox{February 1993}}
\vskip 1.5cm
\centerline{{\bf ANYONIC REALIZATION OF ${\bf SU_q(N)}$ QUANTUM ALGEBRA}}
\vskip 0.6cm
\centerline{{\bf Raffaele Caracciolo}${\ }^\ast$
 ~~and~~ {\bf Marco A. R-Monteiro}${\ }^\ast$
\footnote{$^\dagger $}{Permanent address.}}
\vskip 0.3cm
\centerline{\sl ${\ }^\ast$ Dipartimento di Fisica Teorica}
\centerline{\sl Universit\'a di Torino, and I.N.F.N.
Sezione di Torino}
\centerline{\sl Via P. Giuria 1, I-10125 Torino, Italy}
\vskip 0.35cm
\centerline{\sl ${\ }^\dagger$ CBPF/CNPq, Rua Dr. Xavier Sigaud, 150 }
\centerline{\sl 22290 Rio de Janeiro, RJ, Brazil }
\vskip 2.5cm
\centerline{{\bf Abstract}}
\vskip 0.6cm

\noindent
By considering a set of $N$ anyonic ascillators (non-local, intrinsic
two-dimensional objects interpolating between fermionic and bosonic
oscillators) on a two-dimensional lattice, we realize the $SU_q(N)$
quantum algebra by means of a generalized Schwinger construction. We
find that the deformation parameter $q$ of the algebra is related to
the anyonic statistical parameter $\nu$ by $q=\exp({\rm i}\pi\nu)$.
\bye